%% file: main.tex
\documentclass{ieeeaccess}
\usepackage{cite}
\usepackage{amsmath,amssymb,amsfonts}
\usepackage{algorithmic}
\usepackage{graphicx}
\usepackage{textcomp}
\usepackage{multirow}
\usepackage[linesnumbered,ruled]{algorithm2e}
\usepackage{hyperref}
\usepackage{romannum}
\usepackage{amsmath}
\usepackage{amssymb}
\usepackage{authblk}
\usepackage{float}
\usepackage{commath}
\usepackage{amsfonts}
\usepackage{booktabs}
\usepackage{siunitx}
\usepackage{enumitem}
\usepackage{verbatim}

\def\BibTeX{{\rm B\kern-.05em{\sc i\kern-.025em b}\kern-.08em
    T\kern-.1667em\lower.7ex\hbox{E}\kern-.125emX}}
\begin{document}
\history{Date of publication xxxx 00, 0000, date of current version xxxx 00, 0000.}
\doi{10.1109/ACCESS.2017.DOI}

\title{DeepAoANet: Learning Angle of Arrival from Software Defined Radios with Deep Neural Networks}
\author{\uppercase{Z. Dai}\authorrefmark{1},
\uppercase{Y. He}\authorrefmark{2}, \uppercase{V. Tran}\authorrefmark{3}, \uppercase{N. Trigoni}\authorrefmark{4}, and \uppercase{A. Markham}\authorrefmark{5}}
\address[1,2,3,4,5]{Department of Computer Science, University of Oxford, Oxford, England, United Kingdom (e-mail: $\{$firstname$\}$.$\{$surname$\}$@cs.ox.ac.uk)}

\tfootnote{This research has been financially supported by the National Institute of Standards and Technology (NIST) via the grant Pervasive, Accurate, and Reliable Location-based Services for Emergency Responders (Federal Grant: 70NANB17H185).}

\markboth
{Author \headeretal: Preparation of Papers for IEEE TRANSACTIONS and JOURNALS}
{Author \headeretal: Preparation of Papers for IEEE TRANSACTIONS and JOURNALS}

\corresp{Corresponding author: Z. Dai (e-mail: zhuangzhuang.dai@cs.ox.ac.uk).}

\begin{abstract}
Direction finding and positioning systems based on RF signals are significantly impacted by multipath propagation, particularly in indoor environments. Existing algorithms (e.g MUSIC) perform poorly in resolving Angle of Arrival (AoA) in the presence of multipath or when operating in a weak signal regime. We note that digitally sampled RF frontends allow for the easy analysis of signals, and their delayed components. Low-cost Software-Defined Radio (SDR) modules enable Channel State Information (CSI) extraction across a wide spectrum, motivating the design of an enhanced Angle-of-Arrival (AoA) solution. We propose a Deep Learning approach to deriving AoA from a single snapshot of the SDR multichannel data. We compare and contrast deep-learning based angle classification and regression models, to estimate up to two AoAs accurately. We have implemented the inference engines on different platforms to extract AoAs in real-time, demonstrating the computational tractability of our approach. To demonstrate the utility of our approach we have collected IQ (In-phase and Quadrature components) samples from a four-element Universal Linear Array (ULA) in various Light-of-Sight (LOS) and Non-Line-of-Sight (NLOS) environments, and published the dataset. Our proposed method demonstrates excellent reliability in determining number of impinging signals and realized mean absolute AoA errors less than $2^{\circ}$.
\end{abstract}


\titlepgskip=-15pt

\maketitle

\input{Introduction}

\input{Background}
\input{System_Integration}

\input{Evaluation}

\section{Conclusion}
Extracting AoA of RF signals from passive sensing has enjoyed long-standing research interest due to its importance for a number of applications. Recent SDR front ends make a robust and low-cost AoA estimator possible. However, conventional AoA algorithms fall short in multipath environments with low SNRs. Data-driven AoA methods have emerged in the past decade showing capabilities for extracting features from noisy multichannel data. Yet these methods either require prior knowledge of the signal type, number, and modulation, or merely realize offline performance with synthetic data. We propose DeepAoANet, a pair of Deep Learning models, to determine the AoAs of up to two impinging signals. In doing so, we have collected sample covariance matrices derived from received IQ data in various multipath environments and published the dataset. We have validated the reliability of the DeepAoANet in different scenarios, and found a MAE of less than $2^{\circ}$. This sheds light on a pure physical layer approach to AoA acquisition regardless of the propagation environment, transmission time, or modulation of the signals.

In future work, we shall investigate a finer angular resolution for training with different antenna array setups. We are going to explore how to generalize our approach to signals of other frequency bands, and also in estimating AoA in both azimuth and elevation.

\EOD

\end{document}

%% file: Introduction.tex
\section{Introduction}
\label{sec:intro}

%
%
%
%
\IEEEPARstart{T}{he} modern digital world is constructed in part by the ubiquity of wireless communication, enabling information exchange without mobility constraints. Although the primary purpose is typically for communication, RF signatures in the frequency domain become valuable sources of information in their own right. These signals, also called Signals of Opportunity, have sparked numerous applications such as jammer detection \cite{dai2017}, location fingerprinting \cite{lwu2020}, radio planning etc.

Estimating Angle-of-Arrival (AoA) is one of the most popular radiogoniometric parameters to provide positional awareness about RF signals. Based on their importance, a number of classical algorithms have been proposed to accurately estimate bearing angles, typically based on electromagnetic propagation theory and physical models. One of the most well known is the MUSIC~\cite{music1986} approach for multiple source estimation, with the ability to provide super-resolution beyond the resolution of the antenna array. However, MUSIC performs poorly in NLOS conditions, which is further compounded by operation in low Signal-to-Noise Ratio (SNR) regimes. Moreover, both super-resolution methods and iterative Maximum Likelihood methods require knowing the number of impinging signals \textit{a-priori} \cite{ESPRIT1997, mpastorino2005}.

As an alternative, data-driven or learning-based AoA methods have drawn increasing attention~\cite{zliu2018, myang2020} in the past decade. However, we note that none of the existing data-driven methods utilize real-world channel data for training. Rarely have these methods, which are taught by synthetic data, been validated in real-world environments~\cite{zliu2018}. To some extent, this is due to the high cost and challenges of obtaining basebanded IQ data, as most RF front ends for AoA estimation are dedicated to a specific use case. The high-end USRPs (Universal Software Radio Peripheral) which cover VHF (Very High Frequency), UHF (Ultra High Frequency), and SHF (Super High Frequency) bands are costly, limiting their adoption. We instead focus on investigating how low-cost SDRs, e.g., RTL-SDR dongles, make passive sensing of AoAs possible with a minimal setup. Moreover, state-of-the-art AoA estimator~\cite{obialer2019} which is blind of the impinging signals relies on a 16-element ULA. Yet has it been validated in real-world measurement. To this end, we have proposed a hybrid classification and regression DNN model to predict the AoAs from SDR arrays agnostic of the number of impinging signals or the type of modulation.

We note that AoA is encapsulated through phase-differences between channels due to propagation delays~\cite{akhan2019}. Deep Learning based methods excel at extracting highly-abstract and representative features directly from a large amount of data. This makes Deep Learning an excellent candidate in attending AoA from quadrature demodulated IQ data which are noisy and prone to coupling effect. Regularities of the phase delays can be further revealed by computing correlations among channels, also known as the sample covariance matrix. This acts to reduce the impact of low SNR through correlation over a window of samples. Using the sample covariance matrix to estimate AoA also negates the effect of symbol modulation schemes, pulse shaping, or carrier frequency drift~\cite{jyu2021}. To this end, we have developed a supervised learning approach to AoA using low-cost SDRs. Given a limited number of antenna array elements, our proposed method can estimate multiple AoAs from a single-snapshot sample covariance matrix. However, this task is non-trivial. There are four main difficulties that need to be addressed:

\begin{enumerate}[label=\roman*]
  \item Estimating AoAs in multipath conditions with low SNRs remains a long-standing problem.
  \item Classic AoA methods have a limited Field of View (FOV) \cite{obialer2019} when using a ULA which is prevailing. ULA's (as opposed to Uniform Circular Arrays, UCAs) are commonly encountered due to their slim form-factor.
  \item A blind sensing approach in the physical layer to AoAs is confronted by agnosticism of the signal types, number of sources, or symbol modulation.
  \item Constructing a balanced dataset that allows ergodic supervised-learning is a formidable task because the sheer varieties of multipath channel conditions are infinite.
\end{enumerate}

To the best of our knowledge, state-of-the-art data-driven methods of AoA estimation are all based on training with synthetic data \cite{jyu2021,zliu2018,mgall2020,lwu2019}. Furthermore, real-time performance validation upon multiple signals in multipath environment remains scarce. We utilize a supervised learning approach to predicting up to two AoAs in complex indoor environments. We have taken real-world tests to validate our approach. The source code of this project has been made accessible to the public \footnote{\url{https://github.com/zdai257/DeepAoANet}}. In addition, we have made our dataset accessible to foster research in this area \footnote{\url{https://drive.google.com/drive/folders/1421NOSQcveTE-TpKAM6cPg3SN_vatJPQ?usp=sharing}}.

Our contributions are threefold: (1) We have published the sample covariance matrix feature vectors used for supervised AoA learning, and the corresponding ground-truth AoAs from various real-world environments; (2) We have proposed a series of Deep Neural Network models, termed DeepAoANet, which classify the number of impinging signals before predicting their AoAs accurately without prior knowledge of the signals; (3) Using KerberosSDR, we have developed a real-time AoA estimation and visualization tool to facilitate a low-cost and low-power AoA solution. 

The rest of this paper is organized as follows. Section \Romannum{2} reviews existing AoA estimation methods and advocates the motivations of our approach. The RF front end, DeepAoANet architecture, and dataset composition are expanded in details in Section \Romannum{3}. Performance evaluation of the DeepAoANet is shown in Section \Romannum{4}. Section \Romannum{5} summarizes this work.


%% file: Background.tex
\section{Background}
\label{sec:related_work}

Passive sensing of RF primitives, such as RSSI, TDOA, AoA etc., has seen broad applications in radar, localization, and source detection. RSSI provides a coarse measurement of signal quality versus noise. A limited positioning precision is achievable with RSSI alone \cite{dai2020, abdallah2018}. Localization using TDOA requires rigorous synchronization between transmitter and receivers \cite{zli2018}, which can be expensive and challenging to achieve. AoA on the other hand is a competitive candidate for positioning and source detection in terms of accuracy and cost \cite{rahman2018}.

Listening to RF signals non-intrusively is made possible with low-power and low-cost SDRs. The most remarkable advantage of SDR is the flexibility of adapting to new RF needs based on a common hardware platform \cite{probyns2018}. Compared to the high-end USRP, an RTL-SDR dongle, costing under £20, is a significantly low-cost, low-power, and compact transceiver of VHF and UHF signals \cite{amarquet2020}. The latest RTL-SDR V3 unit produces a maximum sampling rate of $3.2 MHz$ covering frequency bands from $24 MHz$ up to $1.7 GHz$. This gives rise to potentially an extremely low-cost passive AoA estimator in the corresponding frequency bands.

N. BniLam \textit{et al.} \cite{nbnilam2021} proposed an AoA estimator of LoRa signals based on an SDR front end and found $2^{\circ}$ mean error in LOS, and $10^{\circ}$ mean error in NLOS. However, a hand-crafted crystal oscillator has to be integrated to accomplish synchronization among channels. The emerging KerberosSDR \cite{zli2020} comprises four coherent RTL-SDR dongles. The shared crystal oscillator and calibrator board with internal noise source allow sample and phase synchronization with ease. The KerberosSDR makes a cost-effective AoA solution possible.

Conventional AoA algorithms, such as Capon and MUSIC etc., suffer from accuracy degradation in the presence of multipath or a low SNR~\cite{ESPRIT1997, hxiang2019}. Despite widely used with a desirable form factor, the ULA usually results in a limited FOV~\cite{obialer2019}. Taking MUSIC for example, the Wave Vector, $\vec{k}$, propagating along the \textit{x}-axis is defined as

\begin{equation}
\vec{k} = \frac{2\pi}{\lambda} sin\left(\theta\right)
\end{equation}

\noindent
where the wavelength is $\lambda$; the impinging angle with respect to the antenna array is $\theta$. The magnitude of the Wave Vector is Wave Number, $k = 2\pi / \lambda$. Given $M$ array elements, the position of the \textit{m}-th element of the array placed along \textit{y}-axis can be written as;

\begin{equation}
\mathbf{\vec{y}_m} = \left(0, y_m\right)
\end{equation}

\noindent
and

\begin{equation}
y_m = \alpha \left(m - 1\right) \lambda
\end{equation}

\noindent
given the inter-element spacing factor, $\alpha$. The steering vector represents the set of phase delays a plane wave experiences evaluated at a set of array elements, which can be expressed as;

\begin{equation}
\mathbf{\mathbf{a}\left(\theta\right)} = [e^{-j\vec{k}\mathbf{y_1}}, e^{-j\vec{k}\mathbf{y_2}}, \dots, e^{-j\vec{k}\mathbf{y_M}}]^T
\end{equation}

Hence, the received signal, $\mathbf{x}$, can be expressed as;

\begin{equation}
\mathbf{x} = \sum_{l=1}^L \mathbf{a}\left(\theta_l \right) \cdot s_l + \mathbf{n}
\end{equation}

\noindent
where $L$ is the number of impinging signals, $s$ denotes the source signal, and $\mathbf{n}$ is the zero-mean Gaussian noise vector. Given a sampling window of length $N$, the sample covariance matrix, $R$, reads

\begin{equation}
\label{eqn:r}
    \mathbf{R} = \frac{1}{N} \sum\limits_{i=1}^N \mathbf{x} \mathbf{x}^{H}
\end{equation}

\noindent
where $\mathbf{x}^{H}$ denotes the Hermitian Transpose of the received signal.

In order to obtain AoAs of multiple sources, MUSIC takes an eigen-decomposition approach to distinguishing signals from noise which co-exist in a plane wave. The basic hypothesis lies in the orthogonality of signal and noise in the subspace. The eigenvalues and eigenvectors of the signal and noise can be written as;

\begin{equation}
    \mathbf{R} = \mathbf{Q_s} \mathbf{D_s} \mathbf{Q_s}^{H} + \mathbf{Q_n} \mathbf{D_n} \mathbf{Q_n}^{H}
\end{equation}

In ideal conditions, the steering vector in the signal subspace is orthogonal to the noise subspace;

\begin{equation}
\mathbf{\mathbf{a}\left(\theta\right)} \cdot \mathbf{Q_n} = 0
\end{equation}

\noindent
thus,

\begin{equation}
\theta_{MUSIC} = \underset{\theta}{\mathrm{argmin}} \hspace{0.2cm} \mathbf{a}^{H}\left(\theta\right) \mathbf{Q_n} \mathbf{Q_n}^{H} \mathbf{\mathbf{a}\left(\theta\right)}
\end{equation}

Thus, a search is launched over the steering vector candidates to identify the ones that maximize the power spectrum of the received signal;

\begin{equation}
\label{eqn:p}
P\left(\theta\right) = \frac{1}{\mathbf{a}^{H}\left(\theta\right) \mathbf{Q_n} \mathbf{Q_n}^{H} \mathbf{\mathbf{a}\left(\theta\right)}}
\end{equation}

However, MUSIC relies on the orthogonality of signal and noise subspace which does not hold in real-world multipath conditions. Moreover, the model-based derivation of the above power spectrum is susceptible to array positional imperfections and mutual coupling \cite{zliu2018}. To overcome these challenges, the research community have moved towards learning-based AoA in the past few years. Some works attempt to learn AoAs from base-band IQ data directly \cite{akhan2019, mgall2020}. However, received data from RF front ends are subject to various channel characteristics, such as antennas gains, mutual coupling between array elements, negative SNRs etc. Only when the SNR is above $18 dB$ with a fixed window length does the method in \cite{mgall2020} produce acceptable results. In addition, almost all works have used synthetic data for training and validation. It is inevitable that these networks overfit to the assumptions used to simulate the IQ data, which may not hold in the real world.

Learning AoAs from the sample covariance matrix have also drawn increasing attention recently \cite{lwu2019}. The sample covariance matrix calculates correlations among channels while masking the morphology of the signal itself. For instance, SVR (Support Vector Regression) has seen success in generating robust and accurate AoAs from the sample covariance matrix \cite{xhe2010}. SVR is a member of the SVM (Support Vector Machine) family. It shines in mapping multivariate non-linear sample covariance matrix features to latent phase differences. A supervised learning method by applying SVM upon PDPs (Power Delay Profile) of antenna arrays to estimate AoA has been proposed in \cite{myang2020}.

Nevertheless, regression cannot resolve the ambiguity of unknown number of signals at different time instants \cite{wzhu2019}, leading the majority of existing learning approaches to focus on estimating a single source at a time. A two-stage classification model has been proposed in \cite{zliu2018}. In the first stage, a multi-output autoencoder extracts phase delay features from the sample covariance matrix into pre-defined angular subregions. A one-versus-all classifier then determines if a signal exists in each angular subregion. In addition, synthetic noise are introduced in the training data as a form of data augmentation, including gain-phase inconsistency, sensor position errors, and mutual antenna coupling, to increase robustness of the algorithm. However, a fine-grained angular resolution trades off system efficiency and simplicity in constructing the training data, since angular bins of $0.1^{\circ}$ require 1201 classifiers given a FOV = $120^{\circ}$, as well as an immense set of ground-truth labels for training.

In addition, the vast majority of the data-driven methods utilize artificial signals \cite{obialer2019, mgall2020, mcomiter2018}, and few have been validated using real-world signals \cite{zliu2018}. This is probably due to the sheer complexity of real-world RF propagation and signal diversity and the challenges of obtaining sufficient ground-truth data for balanced training. We note that similar issues exist in the acoustic signal processing and speech recognition community. The AoA of acoustic signals have been widely studied using temporal-spectral features from microphone arrays \cite{wqzheng2015, ysun2018}. Despite huge differences in propagating speed and wavelength, acoustic and RF AoA estimation share common challenges such as multipath, also known as reverberation, and interference of multiple sources \cite{mjbianco2020}. S. Adavanne \textit{et al.} \cite{sadavanne2018} proposed a Convolutional Recurrent Neural Network to detect multiple indoor sound sources from the spectrogram of a four-element array. Y. He \textit{et al.} proposed SoundDet \cite{yhe2021} which is an end-to-end sound classification and direction finding system.

In summary, we observe that (a) classical learning approaches, although mature and well studied, do not perform well in high multipath and low SNR conditions, (b) existing learning approaches are predominantly only able to resolve a single source, and (c) more critically, learning approaches are largely trained and validated with simulated data which is not realistic. This leads us to design a learning based system to overcome these challenges.

%% file: System_Integration.tex
\section{DeepAoANet}
\label{sec:system}

We develop a data-driven algorithm to predict AoA's robustly and accurately in multipath conditions. In doing so, we first create a dataset of sample covariance matrix observations in various scenarios. We use relative GPS positions and a compass for direction alignment to label the impinging LoRa signals~\cite{probyns2018}. Thereafter, we generate augmented data by introducing phase shifts and multiple levels of Additive White Gaussian Noise (AWGN) to provide sufficiently dense and balanced training labels~\cite{amelbir2021}. We propose DeepAoANet, a hybrid classification and regression Deep Learning model, to perform end-to-end AoA estimation from a single sample covariance matrix snapshot. It is of key importance for DeepAoANet to learn generic AoA features regardless of signal strength, number of sources, or symbol modulation. We further utilize multi-step fine-tuning techniques~\cite{yshi2019} to jointly train the classification and regression heads. In order to validate the DeepAoANet performance in real-time,  real-world conditions, we have developed a Graphic User Interface (GUI) tool allowing AoA visualization with the aid of KerberosSDR. Moreover, we emphasize that our approach works well with a \textit{single} low-cost SDR device with only 4 antennas. The device can passively sniff the channel, detect any RF transmission (within the frequency band) and estimate the AoA accurately. Such a passive AoA detector is extremely useful in many applications such as indoor localization, intruder detection, etc.

\subsection{Problem Formulation}

Real-world AoA estimation using conventional methods faces numerous challenges. The presence of multipath undermines the assumption that the signal and noise subspace are orthogonal to one another. Coupling between array elements and slight positional imperfections act to reduce the effective FOV. Moreover, it is difficult for a simple peak search algorithm to determine the number of signals from the spatial power spectrum (Eqn. \ref{eqn:p}).

\begin{figure}[!htbp]
	\centering
	\includegraphics[width=0.48\textwidth]{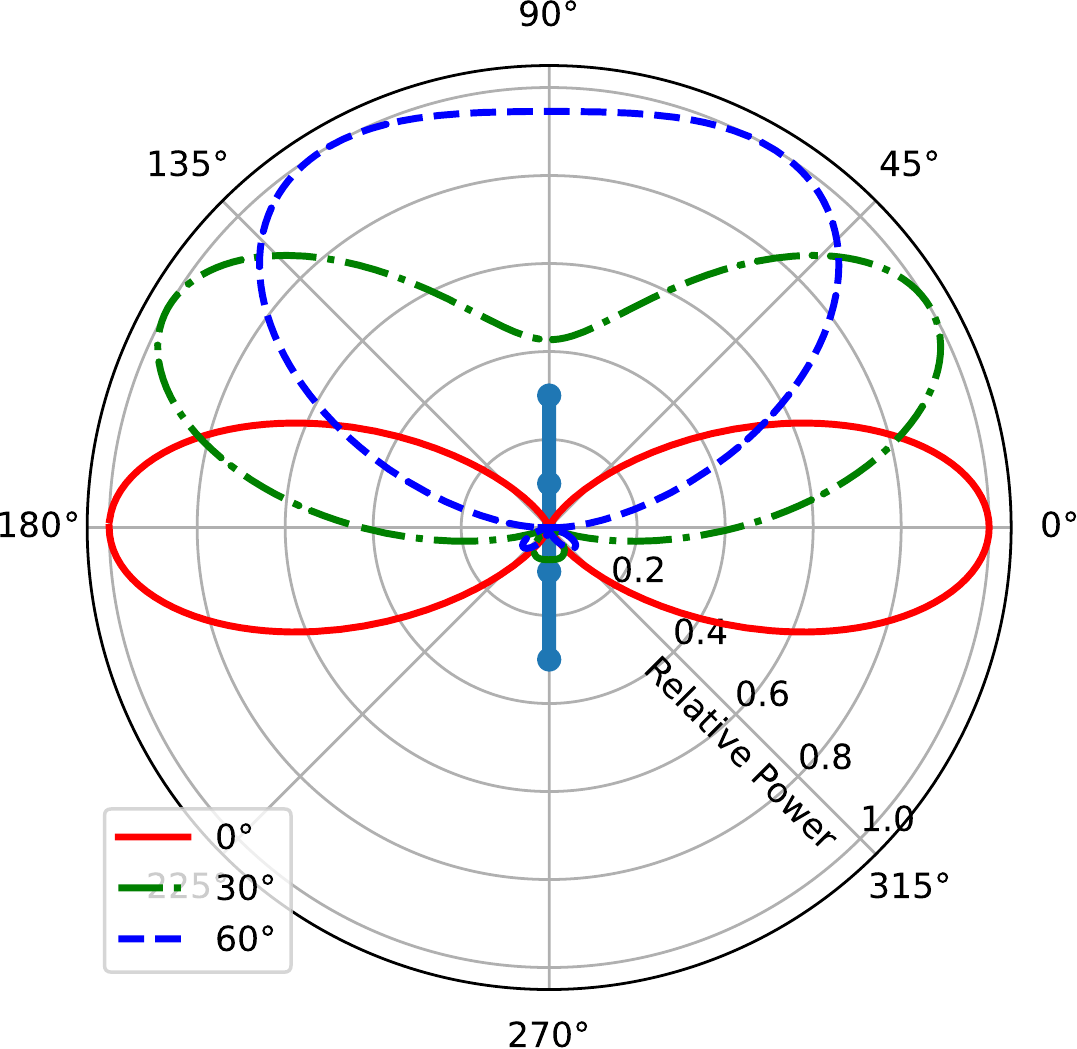}
	\caption[FoVProblem]{Power pattern of a 4-antenna ULA with an inter-element spacing factor ($\alpha$) of $0.25$ with 3 different steering angles of $0^\circ, 30^\circ$ and $60^\circ$ respectively. The beamwidth of antenna array becomes larger when the steering angle moving toward $\pm90^\circ$, and thus more susceptible to noise.}
	\label{ant3angle}
\end{figure}

\begin{figure}[!htbp]
	\centering
	\includegraphics[width=0.48\textwidth]{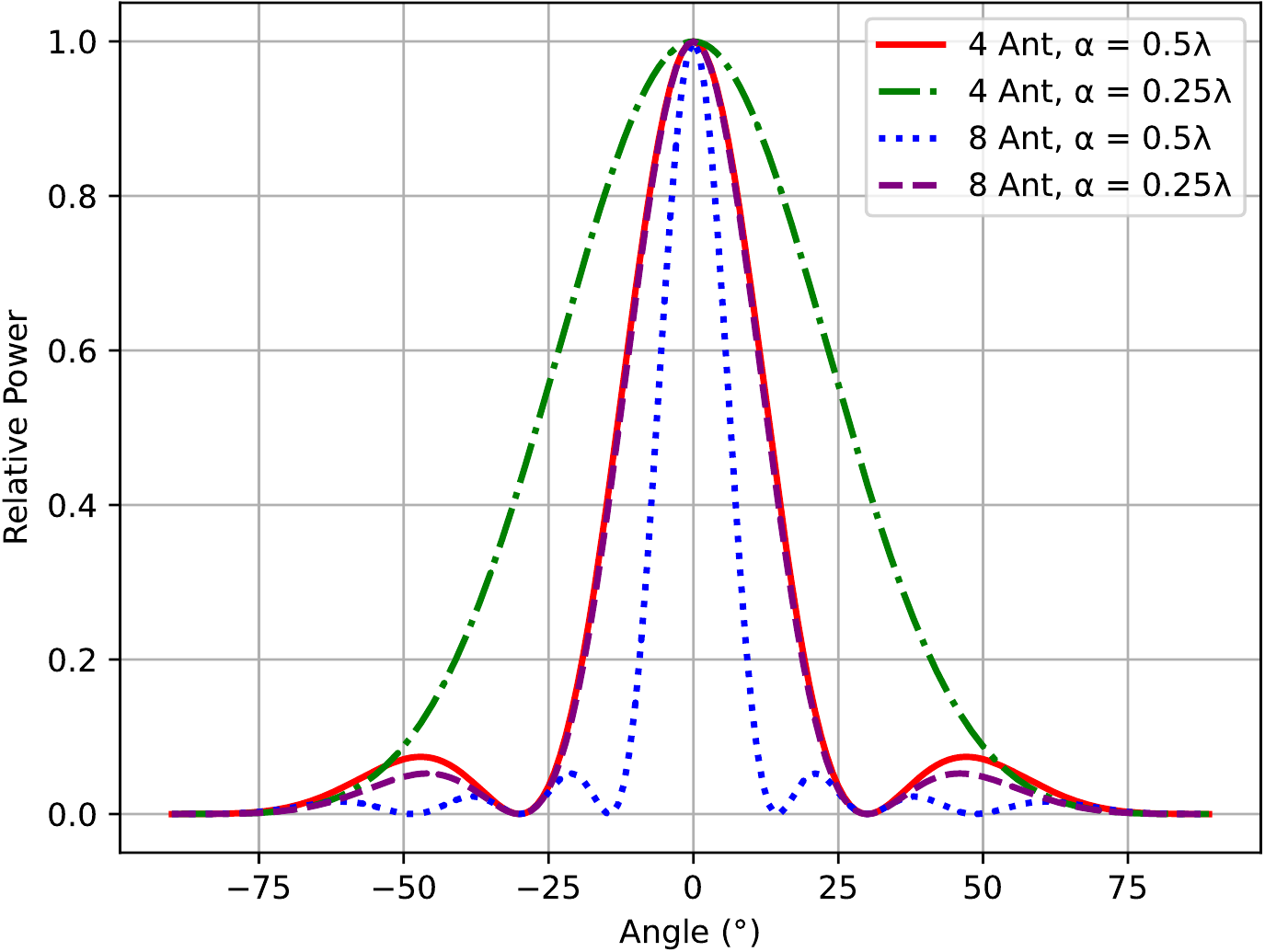}
	\caption[LowSNRProblem]{Power pattern of 4 or 8-antenna ULA configurations (steering angle $= 0^\circ$). A ULA with fewer elements, or with a smaller inter-element spacing factor ($\alpha$), are more susceptible to noise (because of larger beamwidth). However, it may be extremely expensive to accommodate a large number of elements or a large inter-element spacing because of form factor requirements.}
	\label{ant4config}
\end{figure}

Specifically, the first challenge in AoA-based systems is the limited FOV of antenna array (for simplicity, we assume the antenna array is a ULA, although we note that any other geometric structure is applicable). In principle, a ULA is capable of estimating the AoA of arriving signal from $-90^\circ$ to $+90^\circ$. However, the resolution degrades significantly towards the antenna axis (near $+90^\circ$ and $-90^\circ$) and thus limits the effective FOV of the antenna. Figure \ref{ant3angle} plots the power pattern of a 4-element antenna array with an inter-element spacing ($\alpha\lambda$) of $\lambda/4$. The figure shows that the -3dB (half the power) beamwidth of the main beam significantly increases when the steering angle approaches $60^\circ$ (the beamwidth increases from $54^\circ$ to $132^\circ$). Because of the presence of noise, a wider beamwidth will consequently result in higher AoA error.

In real-world environments, robustness to noise is crucial. For AoA-based applications, the size of an antenna array significantly affects the accuracy of AoA estimation in the presence of noise. A smaller antenna array (fewer elements or shorter inter-element spacing) causes wider beamwidth and thus results in lower AoA resolution. Figure \ref{ant4config} shows the power pattern of a ULA with different number of elements (4 and 8) and inter-element spacing ($\lambda/2$ and $\lambda/4$). It is noticeable that the -3dB beamwidth increases more than twice when the number of elements decreases from 8 to 4, or when the inter-element spacing decreases from $\lambda/2$ to $\lambda/4$. Practical indoor systems cannot usually afford an extremely large antenna array for highly accurate AoA estimation. Consequently, a classic highest peak search may fail to detect the correct angle under noisy, multi-path conditions.

Fundamentally, the classic methods are based on a thresholding and/or a searching algorithm to find the most probable impinging signal(s). For example, to find the number of signals, a system can compare the eigenvalues to a threshold; the number of eigenvalues higher than a threshold is considered the number of signals \cite{arraytrack}. However, this method is not reliable when the environment sees a large number of traveling paths and when the antenna array has few elements. As an example (\cite{arraytrack} required 16 and \cite{wiwear} required 8 antennas, with spatial smoothing, to achieve an acceptable AoA accuracy of around $5^\circ$). ArrayTrack\cite{arraytrack} also shows that the direct path is not always the most probable path (in the maximum likelihood sense) in multi-path environments, and thus needs to use multiple antenna arrays to resolve the ambiguity.

In this paper, we explore the use of data-driven approaches to resolve the limitations of classic methods. We argue that a Deep Learning method is able to estimate accurately both the number of impinging signals and the AoA of each individual signal. Given a passive RF receiver system with $M$ synchronized antennas, arranged in some geometric, sub-wavelength spaced configuration. The system passively listens to the RF channel. Whenever the system detects a spike in RF energy, it will execute an algorithm to estimate the number of impinging signals and the corresponding AoA. Instead of relying on a predefined threshold and peak searching algorithms, we design the DeepAoANet that automatically estimates the number of signals and their respective AoA's. In doing so, we have taken a standard method to firstly derive the sample covariance matrix following Eqn. \ref{eqn:r}. The benefits are twofold: Using the sample covariance matrix negates the need to constrain the sampling window to a fixed configuration; Computing the correlations among channels frees the AoA estimation from prior knowledge of signal type or modulation. Therefore, the task of the DeepAoANet is essentially recovering $\theta_l$ from $\mathbf{R}$;

\begin{equation}
\label{eqn:mathproblem}
\theta_1, \theta_2, \dots \theta_L = \textit{g}(\mathbf{R})
\end{equation}

\subsection{DeepAoANet Architecture}

An intuitive solution to derive accurate AoA's is regression. However, classic regression methods, e.g., SVR \cite{mpastorino2005}, cannot compute AoA's directly if the number of signals is unknown. On the other hand, state-of-the-art classification methods require a vast number of angular bins in order to achieve a high angular resolution, as they essentially discretize the angular space. For example, \cite{zliu2018} created 1201 angular bins of $0.1^{\circ}$ which merely covers an FOV of $120^{\circ}$. This not only makes the classification task complex but demands a more sophisticated dataset, e.g., $C_{1201}^2$ classes for labeling. We suggest that this could be a fundamental reason which precludes researchers from using measured data for training.

\begin{figure}[t]
	\centering
	\includegraphics[width=0.48\textwidth]{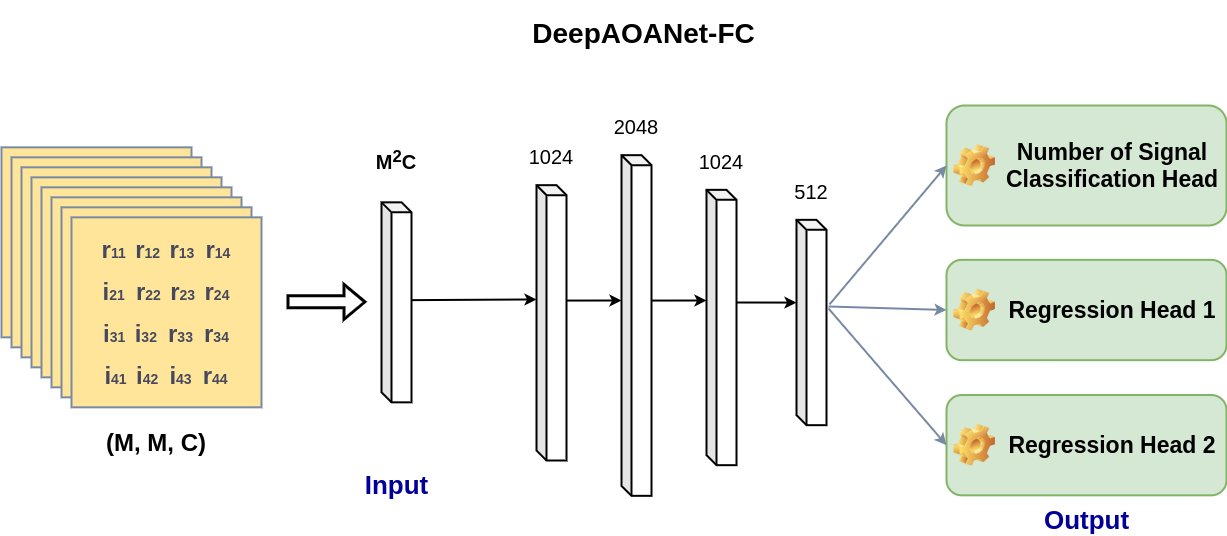}
	\caption[DeepAoANetFC]{Input, output, and the network architectures of the DeepAoANet-FC.}
	\label{DeepAoANetFC}
\end{figure}

Instead, we argue that a hybrid method of joint classification and regression is potentially a more sensible end-to-end model. The intuition behind our approach is that decision on the number of sources is a natural classification problem i.e. choosing from a discrete set of options, whereas estimating angle of arrival is better suited to a regressor operating over a continuous domain. We developed DeepAoANet which comprises a multilayer perceptron\,(MLP), a classfier to estimate the number of impinging signals, and multiple regression heads to predict AoA's. We denote this model as DeepAoANet-FC. The network architecture of the DeepAoANet-FC is shown in Fig.~\ref{DeepAoANetFC}. The layer details of the DeepAoANet-FC are shown in Table~\ref{table:AoAnet1}. Four fully-connected layers with `ReLU' (Rectified Linear Unit) activation are utilized to extract latent phase difference features among the four channels~\cite{zooghby2000}. A classification head predicts the number of sources. Two regression heads predict $\theta_1$ and $\theta_2$, respectively.

\begin{table}[t]
\centering
\caption[DeepAoANet-FC architecture]{DeepAoANet-FC Nerual Network Architecture Illustration. (\textit{B} is batch size; \textit{M} is number of array elements; \textit{C} is number of sample covariance matrix channels).}
\label{table:AoAnet1}
\begin{tabular}{|c|c|c|} 
    \hline
    {Layer Name} & {Units} & {Output Size} \\ 
    \hline
    Input  & n/a  & {$[B, M^2\cdot C]$} \\
    {FC1}  & {$1024$}  & {$[B, 1024]$} \\
    {Dropout1}  & {$rate=0.2$}  & {$[B, 1024]$} \\
    {FC2}  & {$2048$}  & {$[B, 2048]$} \\
    {Dropout2}  & {$rate=0.2$}  & {$[B, 2048]$} \\
    {FC3}  & {$1024$}  & {$[B, 1024]$} \\
    {Dropout3}  & {$rate=0.2$}  & {$[B, 1024]$} \\
    {FC4}  & {$512$}  & {$[B, 512]$} \\
    {Dropout4}  & {$rate=0.2$}  & {$[B, 512]$} \\
    \hline
    {Classification Head} & {Activation=`sigmoid'} & {$[B, 1]$} \\
    {Regression$1$ Head} & {Activation=`sigmoid'} & {$[B, 1]$} \\
    {Regression$2$ Head} & {Activation=`sigmoid'} & {$[B, 1]$} \\ 
    \hline
\end{tabular}
\end{table}

In DeepAoANet-FC, the input feature $input$ is first rearranged into a one-dimensional vector. Then, several learnable fully-connected layers with weight $\mathbf{W}$ and bias term $\mathbf{b}$ individually are applied to sequentially map the input feature to the output feature $output$, with nonlinear activation $ReLU$ between to consecutive fully-connected layers,

\begin{equation}
output = dropout( ReLU(\mathbf{W} \cdot input + \mathbf{b}))
\end{equation}

\noindent
where $dropout(\cdot)$ is a random dropout operator.

\begin{figure}[b]
	\centering
	\includegraphics[width=0.48\textwidth]{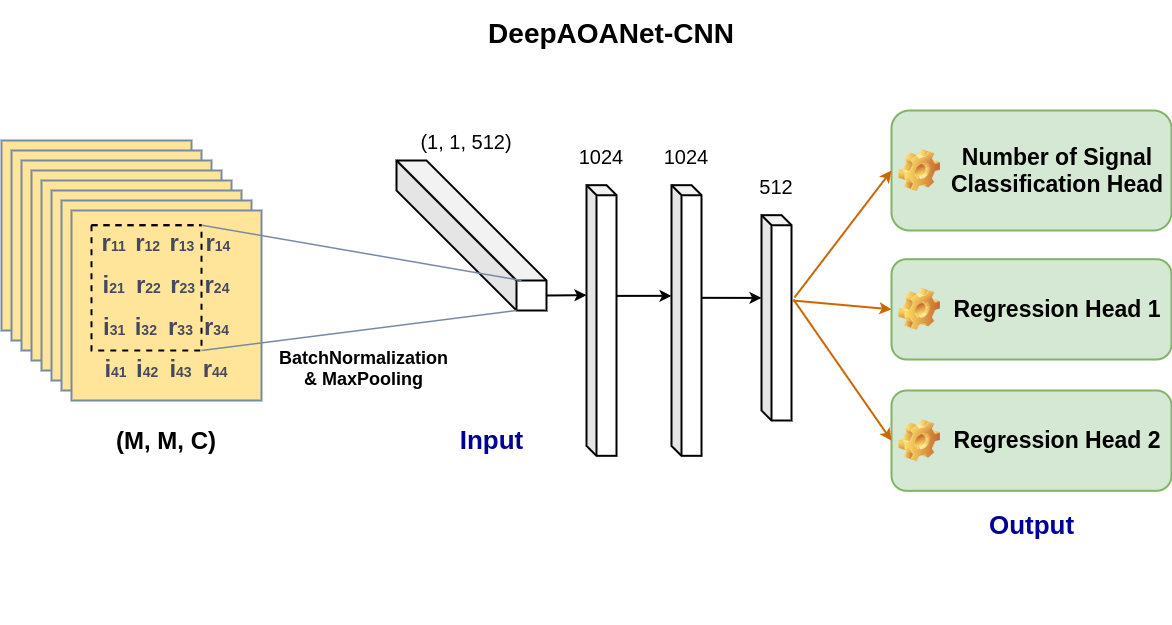}
	\caption[DeepAoANetCNN]{Input, output, and the network architectures of the DeepAoANet-CNN.}
	\label{DeepAoANetCNN}
\end{figure}

We notice the non-diagonal components of the sample covariance matrix are especially small whenever there is no signal. Prevailing acoustic and RF signal processing research all utilize filtering mechanism to determine the presence of a signal~\cite{yhe2021, malygin2017}. This is not only efficient but reliable from our observation because sample covariance matrix essentially computes the correlation of signal co-occurrences among different channels, whereas, noises are independent. Moreover, it is difficult to classify `noise', as it is subject to RF frontend settings and the environment. We argue a predefined threshold can be used to determine whether signals of interest present. Thereafter, if it detects a signal the number of sources and their AoAs can be estimated using DeepAoANet.

The sample covariance matrix is naturally square-shaped in which the latent phase delays reside. Based on this observation, we propose an alternative network architecture using CNN (Convolutional Neural Network) for feature extraction, namely DeepAoANet-CNN. DeepAoANet-CNN takes the input sample covariance matrix as a $M\times M$ `image', and applies a 2D convolution layer before three fully-connected layers. The input `image' can be expanded to $C$ layers by splitting the sampling windows into $C$ smaller windows. This enriches the temporal features of the input by incorporating temporal variations. The network architecture of DeepAoANet-CNN is shown in Fig. \ref{DeepAoANetCNN}.

In DeepAoANet-CNN, the input, $input$, is treated as an image-patch like feature. A 2D convolution $Conv2D$ is applied to consume the image-patch like feature. The same nonlinear operator $ReLU$, as well as random dropout, is applied to the 2D convolved feature before obtaining the output feature $output$,

\begin{equation}
output = dropout( ReLU(Conv2D(input)))
\end{equation}


\noindent
where Conv2D-processed feature is further fed to fully-connected layers for further process.

The network architecture of the DeepAoANet-CNN, as shown in Table \ref{table:AoAnet2}, is similar in which only the first layer is replaced by a 2D Convolution layer with Batch Normailization (BN) and Max-Pooling. A BN layer is inserted right before the first activation function to produce better convergence.

\begin{table}[t]
\centering
\caption[DeepAoANet-FC architecture]{DeepAoANet-CNN Nerual Network Architecture Illustration (\textit{B} indicates the batch size).}
\label{table:AoAnet2}
\begin{tabular}{|ccc|} 
    \hline
    {Layer Name} & {Units} & {Output Size} \\ \hline
    Input  & n/a  & {$[B, M, M, C]$} \\
    {CNN1}  & {$512$, kernel=$3\times3$}  & {$[B, 2, 2, 512]$} \\
    {BN}  & n/a  & {$[B, 2, 2, 512]$} \\
    {Activation}  & {Activation=`ReLU'}  & {$[B, 2, 2, 512]$} \\
    {MaxPool}  & {pool size=$2\times2$}  & {$[B, 1, 1, 512]$} \\
    {FC2}  & {$1024$}  & {$[B, 1024]$} \\
    {Dropout2}  & {$rate=0.2$}  & {$[B, 1024]$} \\
    {FC3}  & {$1024$}  & {$[B, 1024]$} \\
    {Dropout3}  & {$rate=0.2$}  & {$[B, 1024]$} \\
    {FC4}  & {$512$}  & {$[B, 512]$} \\ 
    {Dropout4}  & {$rate=0.2$}  & {$[B, 512]$} \\
    \hline
    {Classification Head} & {Activation=`sigmoid'} & {$[B, 1]$} \\
    {Regression$1$ Head} & {Activation=`sigmoid'} & {$[B, 1]$} \\
    {Regression$2$ Head} & {Activation=`sigmoid'} & {$[B, 1]$} \\ 
    \hline
\end{tabular}
\end{table}

Since the classification and regression heads use different loss metrics, namely, Binary Cross-Entropy (BCE) and Mean-Squared Error (MSE), a loss weight coordination is necessary in the joint training process. The BCE loss function can be written as;

\begin{equation}
\mathbf{L_c} = - \frac{1}{N} \sum_{i=1}^N \left[ y_i \log(p_i) + (1-y_i) \log(1 - p_i) \right]
\end{equation}

\noindent
where $y_i$ is the ground-truth label of having two sources for the $i$-th sample, and $p_i$ is the corresponding probability from the `sigmoid' activation. On the other hand, either regression loss can be written as;

\begin{equation}
\mathbf{L_r}(z, \hat{z}) = \frac{1}{N} \sum_{i=1}^N \left( z_i - \hat{z}_i \right)^2
\end{equation}

\noindent
where $z$ is the ground-truth angle and $\hat{z}$ is the prediction. We define the joint loss function as following;

\begin{equation}
\label{eqn:loss}
\mathbf{L} = \tau \cdot \left[\mathbf{L_c}, \mathbf{L_r}(z_1, \hat{z}_1), \mathbf{L_r}(z_2, \hat{z}_2) \right]^T
\end{equation}

Notice \cite{obialer2019} only considers correctly classified numbers of signal in the mean error calculations which overestimates its realistic performance. We have defined the RMSE and MAE metrics, with penalties for incorrect classifications as follows;

\begin{equation}
RMSE = 
\begin{cases}
\sqrt{\frac{\sum_{i=1}^N \left[ \frac{(\hat{\theta}_{i1} + \hat{\theta}_{i2})}{2} - \theta_{i1} \right]^2}{N} } &\text{if $L=1$}\\
\sqrt{\frac{\sum_{n=1}^N \left[ (\hat{\theta}_{i1} - \theta_{i1})^2 + (\hat{\theta}_{i2} - \theta_{i2})^2 \right]}{N}} &\text{if $L=2$}
\end{cases}
\end{equation}

\begin{equation}
MAE = 
\begin{cases}
\frac{\sum_{i=1}^N \| \frac{(\hat{\theta}_{i1} + \hat{\theta}_{i2})}{2} - \theta_{i1} \|}{N}  &\text{if $L=1$}\\
\frac{\sum_{n=1}^N \|\hat{\theta}_{i1} - \theta_{i1} \| + \|\hat{\theta}_{i2} - \theta_{i2} \|}{N} &\text{if $L=2$}
\end{cases}
\end{equation}

\subsection{Dataset Composition}
\label{subsec:sdr}

To conduct supervised learning upon DeepAoANet, a crucial step is to construct a balanced and representative dataset. We used intermittent LoRa beacons at $0.5 Hz$ as the vendor signal, which transmits a narrow-band signal at $868 MHz$ in the EU. The benefits are threefold: (1) excellent range in multipath environments; (2) highly configurable air time, transmitting power, and payload \cite{mahfoudi2019}; (3) free licensing in the Industrial, Scientific and Medical (ISM) band.

\begin{figure}[t]
	\centering
	\includegraphics[width=0.47\textwidth]{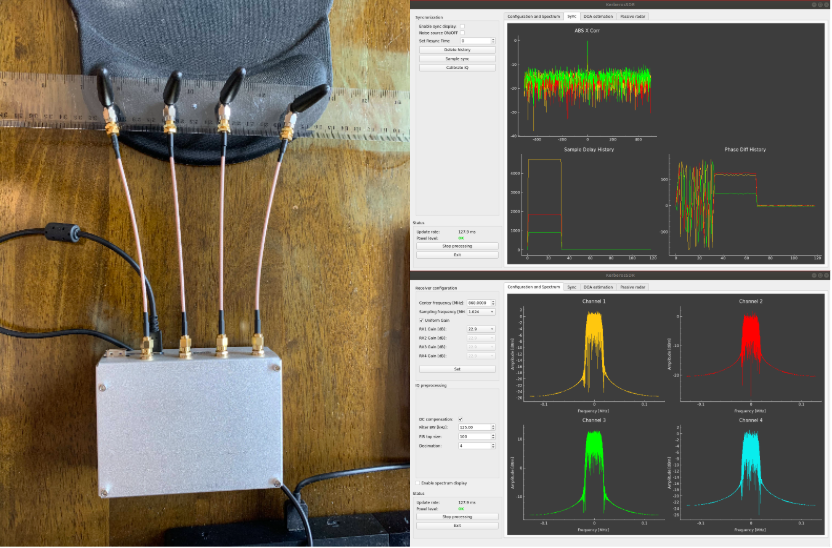}
	\caption[KerberosSDR]{On the left is a KerberosSDR connected to four whip antennas with an inter-element spacing factor $\alpha = 0.2$. The up-right screenshot shows the sample and phase synchronization step. The bottom-right shows the spectrum of four channels after band-pass filtering.}
	\label{KerberosSDR}
\end{figure}

We used KerberosSDR as the receiver for data collection thanks to its low cost and simplified synchronization steps. An open-source driver of KerberosSDR with GUI is provided. The KerberosSDR driver enables real-time AoA visualization in several steps as shown in Fig. \ref{KerberosSDR}. Several conventional AoA algorithms are built-in such as Capon \cite{capon1995} and MUSIC. We took the following procedures to setup the KerberosSDR as a passive LoRa receiver;

\begin{description}
  \item[$\bullet$] Connect KerberosSDR to a laptop through USB and launch the driver.
  \item[$\bullet$] Specify the center frequency as $868 MHz$, sampling rate (
  $2 MHz$), a unform gain for four channels ($22.9 dB)$, and bandwidth ($125 kHz$).
  \item[$\bullet$] Disconnect the antennas. Activate the noise source before conducting sample and phase synchronization.
  \item[$\bullet$] Connect the antennas. Apply a band-pass filter, with FIR tap size 100 and decimation 4, at the center frequency.
  \item[$\bullet$] Specify the ULA inter-element spacing factor as 0.2. Activate AoA estimation.
\end{description}

We labeled the data using relative GPS positions and a compass for direction alignment. A Dragino LoRa-GPS board stacked on a Raspberry Pi 4 was used as the transmitter. The LoRa payload contains a full NMEA Message \cite{dai2020}. We have specified the LoRa SF (Spreading Factor) equal to 11. This not only increases the range but lengthen the air time to increase the efficiency of signal recording.

During data collection, we collected sample covariance matrix snapshots measured in a variety of propagation conditions. Nearly 50k snapshots of sample covariance matrices were collected in (a) open area, (b) open area with copper sheet as reflectors, (c) office, (d) corridor, and (e) interior-to-outdoor environments. The composition of the training dataset is illustrated in Table \ref{table:data}.

\begin{table*}[!htpb]
\centering
\caption[Training Data Composition]{Training Data Composition}
\label{table:data}
\begin{tabular}{|c|c|c|c|c|c|}
\hline
\bfseries{Dataset No.} & Data Collection Scenario & Raw & Phase-shifted & Synthetic AWGN or Carrier Phase & Label\\ \hline
\bfseries{1} & LOS & 8.8k & 35.5k & 133.3k & \multirow{5}{*}{$(Class_{L=1}, \theta_1, \theta_1)$}\\ \cline{1-5}
\bfseries{2} & LOS with rotating copper reflectors & 10.1k & 40.2k & 150.8k & \\ \cline{1-5}
\bfseries{3} & NLOS office environment & 11.5k & 46.1k & 172.8k & \\ \cline{1-5}
\bfseries{4} & NLOS corridor environment & 9.1k & 36.3k & 136k & \\ \cline{1-5}
\bfseries{5} & NLOS interior-to-outdoor & 7.2k & 28.8k & 108.1k & \\ \hline
\bfseries{6} & Combination of \bfseries{[2, 3]} & 29k & 164.3k & 164.3k & \multirow{6}{*}{$(Class_{L=2}, \theta_1, \theta_2)$}\\ \cline{1-5}
\bfseries{7} & Combination of \bfseries{[2, 4]} & 27.6k & 156.5k & 156.5k & \\ \cline{1-5}
\bfseries{8} & Combination of \bfseries{[2, 5]} & 23.3k & 132.3k & 132.3k & \\ \cline{1-5}
\bfseries{9} & Combination of \bfseries{[3, 4]} & 24.2k & 137.2k & 137.2k & \\ \cline{1-5}
\bfseries{10} & Combination of \bfseries{[3, 5]} & 19k & 108k & 108k & \\ \cline{1-5}
\bfseries{11} & Combination of \bfseries{[4, 5]} & 19.8k & 112.1k & 112.1k & \\ \hline
\end{tabular}
\end{table*}

\begin{figure}[t]
	\centering
	\includegraphics[width=0.37\textwidth]{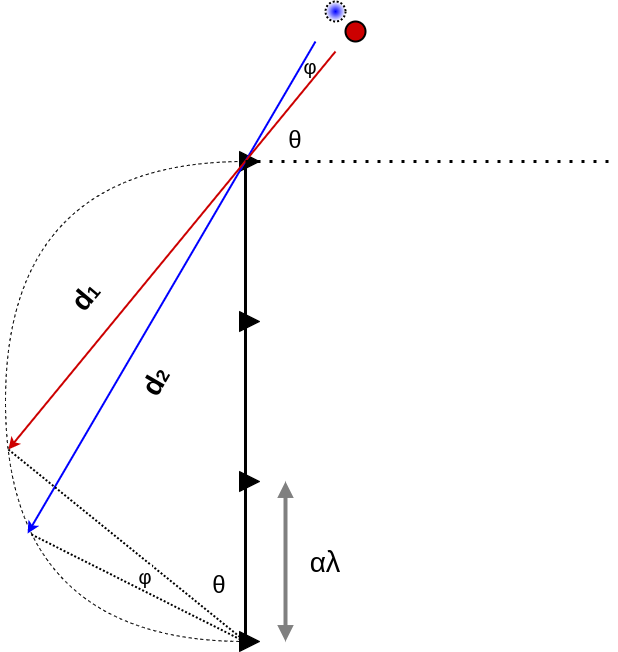}
	\caption[phasegeometry]{Effects of impinging angle change, $\varphi$, to the delay of the last channel with respect to the first channel. The red circle represents original source location while the red quiver represents the impinging angle, $\theta$; Blue quiver stands for the new impinging angle after a phase augmentation where the blue circle is the virtual source; Black triangles are the ULA elements with a spacing factor of $\alpha$. The phase delay (or advancement), $\triangle\xi$, is equal to $(d_2 - d_1) / \lambda$. Note that smaller impinging angles see greater phase shifts given the same $\varphi$.}
	\label{phasegeometry}
\end{figure}

Discrete ground-truth AoA labels, between $[-70^{\circ}, 70^{\circ}]$ with $10^{\circ}$ increment, were recorded by referring to the GPS positions. We observed a threshold can be used upon $R_{ij}$ values to differentiate signals from noise. If over $90\%$ of non-diagonal $R_{ij}$ magnitudes are above $1e^{-4}$, signals are detected. For each category of environment, Additive White Gaussian Noises (AWGN) of zero mean and different Standard Deviations have been introduced to enrich the dataset. This emulates receiving signals at different SNRs or ranges.

To approach an enhanced angular resolution, we have introduced phase shifts to the dataset. A discrete angular increment or decrement of $2^{\circ}$ is posed to the original data as illustrated in Fig. \ref{phasegeometry}. This angular augmentation is restricted within $4^{\circ}$ to best preserve the ratio of raw data versus synthetic and enrich the labels for regression. Therefore, the $m$-th array element of the receiver sees a phase shift as below

\begin{equation}
\mathbf{\hat{x}} = \mathbf{x} \cdot e^{j\triangle\xi} = \mathbf{x} \cdot e^{j(m-1)\alpha[\sin{\theta} - \sin{(\theta + \varphi)}]}
\end{equation}

\noindent
where $\theta$ is the ground-truth AoA and $\varphi \in \{-4^{\circ}, -2^{\circ}, 2^{\circ}, 4^{\circ}\}$. Hence, a FOV ranging from $-74^{\circ}$ to $74^{\circ}$ has been created for a fine-grained AoA learning.

Similarly, AoA detection of multiple sources are constructed through superposition of the collected single-source data. The IQ data collected at scenarios other than LOS of different ground-truth AoAs are superimposed to create synthetic IQs. Notice the carrier signals come with stochastic initial phases wherever they interfere. A random phase difference between the carriers is inserted to simulate constructive or destructive interference. The steps are expanded as follows.

The modulated signals at carrier frequency, $F_0=868 MHz$, can be written as;

\begin{equation}
\mathbf{m_1}(t) = x_1(t) e^{j\phi_1} e^{j2\pi F_0 t} = \left[ i_1(t) + jq_1(t) \right] e^{j2\pi F_0 t}
\end{equation}

\begin{equation}
\mathbf{m_2}(t) = x_2(t) e^{j\phi_2} e^{j2\pi F_0 t} = \left[ i_2(t) + jq_2(t) \right] e^{j2\pi F_0 t}
\end{equation}

\noindent
where $x_1(t)$ and $x_2(t)$ are the base-band signals, and $\phi_1$ and $\phi_2$ stand for the initial phases of the carrier signals which are independent. Only in the modulated frequency band are the signals additive. As a result, the superimposed signal reads

\begin{equation}
\mathbf{m_1}(t) + \mathbf{m_2}(t) = \left[ x_1(t) + x_2(t) e^{j(\phi_2-\phi_1)} \right] e^{j\phi_1} e^{j2\pi F_0 t}
\end{equation}

If we define the carrier phase difference $\triangle \phi = \phi_2 - \phi_1$, IQ samples of the superimposed signal with random original carrier phases yields

\begin{equation}
i_1(t) + jq_1(t) + \left[ i_2(t) + jq_2(t) \right] e^{j\triangle \phi} \hspace{1.5em} \triangle \phi \in \left[ 0, 2\pi \right)
\end{equation}

\noindent
in which $\triangle \phi$ conforms continuous uniform distribution within a period.

The phase differences of the SDR channels lie in the upper triangular components of the sample covariance matrix, $R$, as values of the upper-right and lower-left triangles are symmetric. The diagonal components are always real numbers equal to variances of the received signals. Therefore, we serialized the real parts of $R$ in the top-right triangle, together with imaginary parts of the bottom-left triangle, as the input feature vectors;

\begin{equation}
\mathbf{b_r} = \left[ r_{11}, r_{12}, \dots, r_{jk}, \dots, r_{44}\right]^T \hspace{2em} \text{for $j \leq k$}
\end{equation}

\begin{equation}
\mathbf{b_i} = \left[ i_{21}, i_{31}, \dots, i_{jk}, \dots, i_{43}\right]^T \hspace{2.2em} \text{for $j > k$}
\end{equation}

\begin{equation}
\label{eqn:b}
\mathbf{b} = \left[ \mathbf{b_r}; \mathbf{b_i} \right]
\end{equation}

\begin{equation}
\label{eqn:input}
input = \frac{\mathbf{b}}{\| \mathbf{b} \|}
\end{equation}

The KerberosSDR sampling window length is configured as $2^{15}$. In this work, we chose a sliding window of length $2^{12}$ to generate an 8-window sample covariance matrix. This not only exploits temporal variations but enriches the input feature space. Since there are four array elements in KerberosSDR, each sample covariance matrix channel comprises $M \times M = 16$ floating-point numbers. This gives a 128-element feature vector, or a feature `image' of size $(4, 4, 8)$. We utilized standard scaling to normalize all the inputs.

\begin{figure*}[!htbp]
	\centering
	\includegraphics[width=0.87\textwidth]{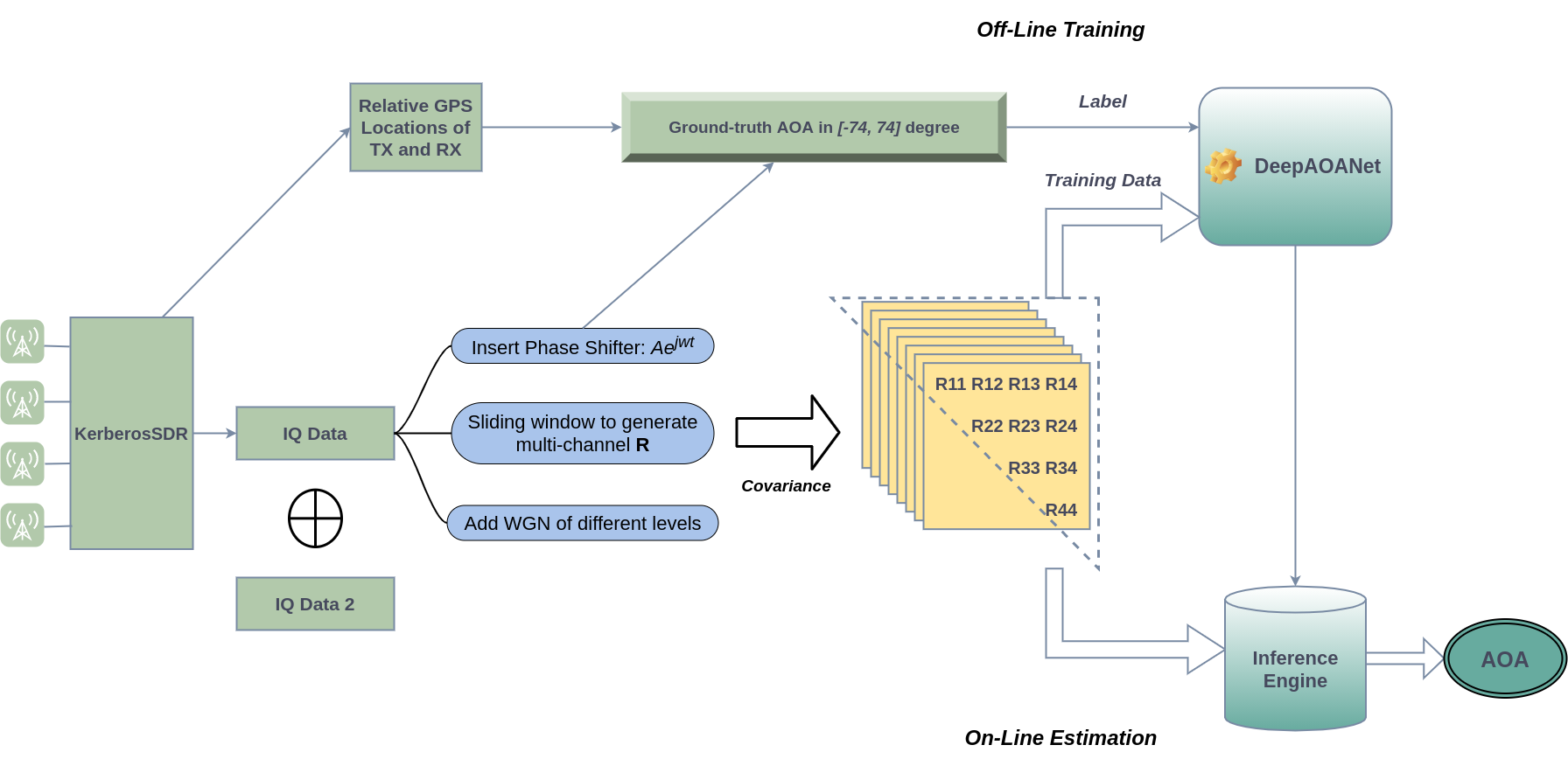}
	\caption[Flowchart]{A flowchart demonstrating how the offline training and online AoA estimation work.}
	\label{Flowchart}
\end{figure*}

As for labels, we have arranged them as $(Class_L, \theta_1, \theta_2)$, in which $\theta_1$ is populated if only one source exists. In this work, we only consider three circumstances as $L \in \{0, 1, 2\}$. Since no-signal case can be solved effectively with a threshold-based detector, we assigned binary classification of $L=1$ and $L=2$ for the classifier. As ground-truth AoAs fall in $\left[-74^{\circ}, 74^{\circ} \right]$, we uniformly projected $\theta$ to be within $(0, 1)$. Note that $\theta_1$ and $\theta_2$ are rearranged in ascending order to avoid confusing the regression heads.

We utilized ROS (Robot Operating System) for field data collection as well as data synthesis. The total number of sample covariance matrix snapshots exceed 2.6 million. The process of data collection, labeling, and training are outlined in Fig. \ref{Flowchart}. We have further validated that our dataset contains sufficient amount of training sources as well as label diversity in Section \ref{sec:eval}.

\subsection{AoA Inference and Visualization}

\begin{figure}[!htbp]
	\centering
	\includegraphics[width=0.46\textwidth]{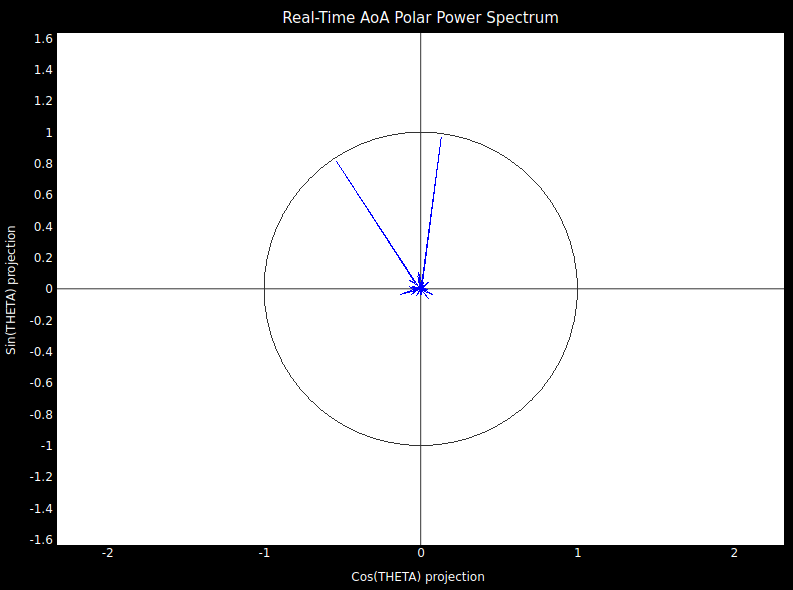}
	\caption[GUI2sig]{An example of visualizing two impinging signals from $-30^{\circ}$ and $10^{\circ}$ using the developed GUI tool.}
	\label{GUI2sig}
\end{figure}

In order to validate our approach, we built an inference engine of the DeepAoANet for real-time estimation. A GUI is developed to visualize the impinging signals. At runtime, two LoRa sources are configured with different SF and power settings transmitting messages of random lengths. To test the generalization capacity of the DeepAoANet, the transmitters have been placed in different indoor and outdoor environments. An example of the GUI showing two impinging signals are displayed in Fig. \ref{GUI2sig}. A detailed performance analysis is presented in Section \ref{sec:eval}.

%% file: Evaluation.tex
\section{Evaluation}
\label{sec:eval}

We have trained and tested the DeepAoANet using Keras with Tensorflow backend. Adam optimizer with an initial learning rate of $1e^{-3}$ is used with a $1e^{-6}$ decaying factor. We modify the weight vector $\tau$ in the loss function (Eqn.~\ref{eqn:loss}) in a two-stage training process: for the first 40 epochs, $\tau$ equals $[0.1, 1, 1]$; for the last 10 epochs, $\tau = [0.001, 1, 1]$ is assigned. Because the BCE loss stops declining at a relatively large value with classification saturates above $99\%$ accuracy, we suppress its weight in the loss to allow regressions proceed further. We split our dataset into $60\%$ for training, $30\%$ for validation, and $10\%$ for testing, with a batch size of 512. It takes 1.5 hours and 2 hours on a Dell XPS 15 laptop with an Intel Core i7 CPU and a GeForce GTX 1650 GPU to train DeepAoANet-FC and DeepAoANet-CNN, respectively.

To validate the effectiveness of our proposed DeepAoANet, we contrast DeepAoANet's against other classic and regression methods. We have conducted tests in unseen environments with different LoRa configurations under various SNRs. We have investigated the error distributions of DeepAoANet to validate its effetive FOV. We have measured the DeepAoANet's latency on different computing platforms to validate its efficiency.

\subsubsection{Comparison to other methods}

\begin{figure}[!htbp]
	\centering
	\includegraphics[width=0.46\textwidth]{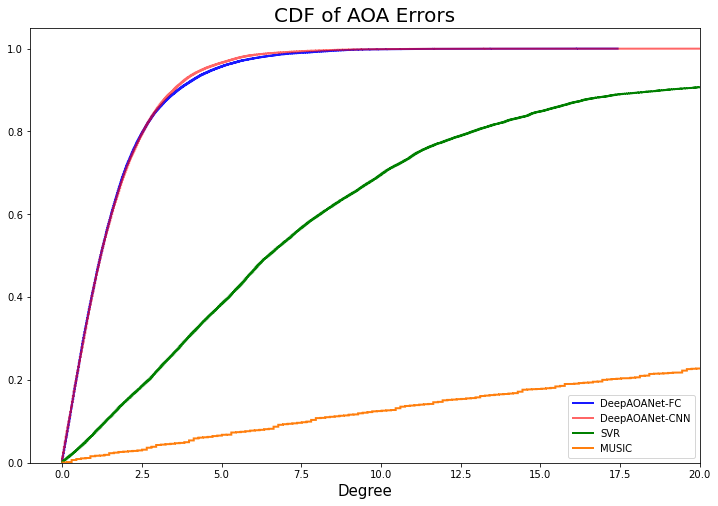}
	\caption[CDF]{CDF of estimating single-source AoA. Notice MUSIC is severely under-performing because its reliable FOV stays between $(-45^{\circ}, 45^{\circ})$ from our observation. Whereas, the ground-truth angles in this benchmark span across $(-74^{\circ}, 74^{\circ})$.}
	\label{CDF}
\end{figure}

Since SVR can only deal with single source~\cite{xhe2010} and MUSIC performance is subject to the selected peak search algorithm, we evaluated DeepAoANet against other methods using only the single-source split with a $0 dB$ SNR for a fair comparison in the Cumulative Distribution Function (CDF) plot. The CDFs of DeepAoANet-FC, DeepAoANet-CNN, MUSIC, and SVR are displayed in Fig.~\ref{CDF}. As can be seen, both DeepAoANet candidates yield RMSEs of approximately $2.5^{\circ}$ for $80\%$ of the tested samples. In contrast, SVR produces a RMSE of more than $10^{\circ}$. Notice MUSIC fails in this benchmark due to its limited FOV when estimating large impinging angles.

\subsubsection{Effective FOV}

\begin{figure*}[!htbp]
	\centering
	\includegraphics[width=0.92\textwidth]{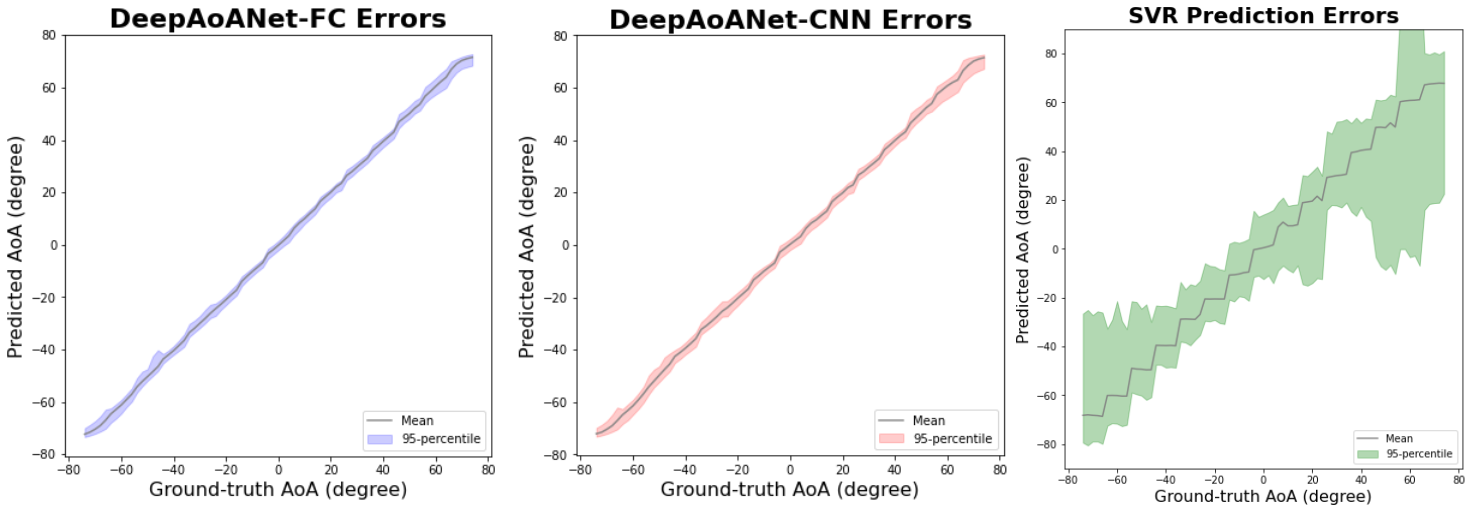}
	\caption[scatter]{Distributions of 95-percentile most accurate AoA's predictions using DeepAoANet-FC, DeepAoANet-CNN, and SVR.}
	\label{scatter}
\end{figure*}

We have investigated where the errors reside in the angular spectrum using the DeepAoANet candidates versus SVR as shown in Fig.~\ref{scatter}. Generally, the errors are evenly distributed. To quantify this statement, we calculated the error standard deviations when the ground-truth AoAs are within $(-37^{\circ}, 37^{\circ})$ or outside. They are $2.101^{\circ}$ and $2.454^{\circ}$ using DeepAoANet-FC, or $1.954^{\circ}$ and $2.369^{\circ}$ using DeepAoANet-CNN, respectively. It can be seen larger impinging angles are more difficult to estimate. In contrast, SVR suffers large estimation errors with large impinging angles. The standard deviation for AoA's within $(-37^{\circ}, 37^{\circ})$ equals $9.202^{\circ}$; for AoA's outside $(-37^{\circ}, 37^{\circ})$ it is $12.845^{\circ}$. Nevertheless, DeepAoANets have demonstrated a superb stability across a wider FOV in comparison to SVR.

\subsubsection{Accuracy versus SNRs}

\begin{figure}[!htbp]
	\centering
	\includegraphics[width=0.47\textwidth]{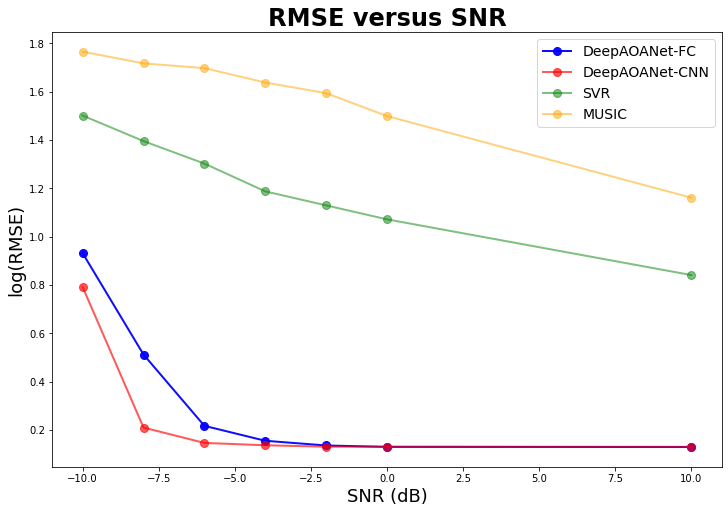}
	\caption[snr]{Logarithm of RMSE to base $10$ of the DeepAoANets, SVR, and MUSIC against varying SNRs. Note SVR is analysed upon single-source split of the testing dataset.}
	\label{snr}
\end{figure}

It can be seen from Fig.~\ref{snr} both DeepAoANet-FC and DeepAoANet-CNN exhibit great resilience dealing with negative SNRs i.e. signals below the noise floor. Either SVR or MUSIC performs poorly. Notice DeepAoANet-CNN demonstrates better reliability when the SNR is less than $-5 dB$. We think this owes to the DeepAoANet-CNN's superb power in generalizing to deteriorated SNRs that are lower than what is presented during training.

We investigated the classification performance when the SNR equals $0 dB$. The classification confusion matrices of the DeepAoANet-FC and DeepAoANet-CNN are shown in Table~\ref{table:c_mtx_fc} and Table~\ref{table:c_mtx_cnn}. The DeepAoANet candidates demonstrate excellent capabilities in identifying the number of impinging signals.


\begin{table}[!htpb]
\centering
\caption[DeepAoANet-FC cmtx]{Confusion Matrix of DeepAoANet-FC Classifier}
\label{table:c_mtx_fc}
\begin{tabular}{|c|c|c|c|}
\hline
\multicolumn{2}{|c|}{Accuracy ($\%$)}&\multicolumn{2}{c|}{True Number of Signals}\\
\cline{3-4}
\multicolumn{2}{|c|}{ } &$L=1$&$L=2$\\
\hline
\multirow{2}{*}{Prediction}& $L=1$ & $100.0\%$ & $0.151\%$ \\
\cline{2-4}
& $L=2$ & $0.0$ & $99.849\%$ \\
\hline
\multicolumn{2}{|c|}{Total Samples} & 96635 & 175764\\
\hline
\end{tabular}
\end{table}

\begin{table}[!htpb]
\centering
\caption[DeepAoANet-CNN cmtx]{Confusion Matrix of DeepAoANet-CNN Classifier}
\label{table:c_mtx_cnn}
\begin{tabular}{|c|c|c|c|}
\hline
\multicolumn{2}{|c|}{Accuracy ($\%$)}&\multicolumn{2}{c|}{True Number of Signals}\\
\cline{3-4}
\multicolumn{2}{|c|}{ } &$L=1$&$L=2$\\
\hline
\multirow{2}{*}{Prediction}& $L=1$ & $100.0\%$ & $0.22\%$ \\
\cline{2-4}
& $L=2$ & $0.0$ & $99.78\%$ \\
\hline
\multicolumn{2}{|c|}{Total Samples} & 96635 & 175764\\
\hline
\end{tabular}
\end{table}

\subsubsection{Real-world test}

\begin{figure}[!htbp]
	\centering
	\includegraphics[width=0.47\textwidth]{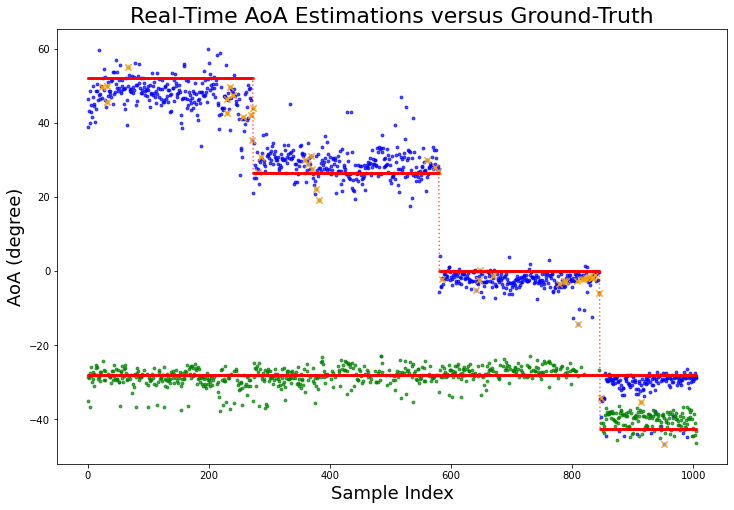}
	\caption[qualitative]{Real-time AoA estimations versus the ground-truth using DeepAoANet-CNN. The red lines mark the ground-truth AoAs of the two LoRa sources. The green dots represent the 1st regression output. The blue dots represent the 2nd regression output. The orange crosses stand for samples with wrong classification outcomes.}
	\label{qualitative}
\end{figure}

To verify DeepAoANet can handle different types of signals, we have placed the KerberosSDR and two LoRa transmitters inside a laboratory with a Vicon system to provide ground-truth device positions of a $mm$-level precision, so as to ground-truth AoAs. The laboratory is in its typical state filled with items. The KerberosSDR remains static while one LoRa source is configured with SF 9 and $17\hspace{0.1cm}dBm$ of transmitting power, and the other at SF 12 and $10\hspace{0.1cm}dBm$ of transmitting power. Both were generating payloads of random lengths. During the test, one transmitter was fixed and the other was placed at four different locations. The estimated AoAs using DeepAoANet-CNN and the ground-truths are plotted in Fig.~\ref{qualitative}. This proves the DeepAoANet generalizes well in unseen environments dealing with signals without knowing the symbol modulation.

\subsubsection{Run-time efficiency}

Both DeepAoANet-FC and DeepAoANet-CNN have merely a few million parameters. This makes them suitable for real-time applications. We have logged the DeepAoANets', as well as MUSIC and SVR's, runtime performance on platforms with different resource constraints, including the Dell XPS15 laptop, an NVIDIA Jetson AGX Xavier with a 512-core Volta GPU, and a Raspberry Pi 4, as shown in Table~\ref{table:runtime}. It can be seen using DeepAoANet inference is more efficient compared to MUSIC (based on \textit{numpy} library). SVR is significantly faster. Nonetheless, the DeepAoANets allow inference rates over 60 samples-per-second even with a single-board computer.

\begin{table}[!htbp]
\centering
\caption[Inference runtime benchmark]{Inference latency benchmark}
\label{table:runtime}
\begin{tabular}{|c|c|c|c|c|}
\hline
\multirow{2}{*}{\bfseries{Model}} & \multirow{2}{*}{Parameters} & \multicolumn{3}{|c|}{Latency (ms)}\\ \cline{3-5}
 &  & Laptop & Jetson & Pi 4\\ \hline
MUSIC & n/a & 5.973 & 19.069 & 61.277\\ \hline
SVR & n/a & 0.869 & 2.071 & 5.721\\ \hline
DeepAoANet-FC & 4.86M & 5.988 & 4.829 & 15.212\\ \hline
DeepAoANet-CNN & 2.14M & 4.436 & 5.116 & 7.984\\ \hline
\end{tabular}
\end{table}

\subsubsection{Limitations}

There are several limitations in our current approach. The DeepAoANet can only handle up to two sources at the moment. This is because the amount of training data becomes exponentially larger with the number of sources. Given 75 angular regions for regression, there are $C_{75}^3 = 67525$ combinations of labels, each of which needs to be contaminated by different levels of noise. We plan to use semi-supervised learning rather than ergodic supervised learning to tackle more than two AoA's in future work. Although both configurations of DeepAoANet delivered good results dealing with LoRa signals of arbitrary configurations, they failed when we shifted the transmitter center frequency by $100\hspace{0.15cm}kHz$. The accuracy also severely degraded if we changed the antenna inter-element spacing factor at runtime. As specified in Section \ref{subsec:sdr}, only when the KerberosSDR is tuned to the same center frequency, bandwidth, sampling rate, and decimation does the DeepAoANet find its ground for phase delay analysis. This is because the IQ data is subject to the RF front end setup in which the training dataset were collected. In the next step, we will collect more data under different setups in pursuit of a truly generic AoA solution independent of the RF front end or carrier frequency.

%% file: main.bbl
\begin{thebibliography}{1}

\bibitem{dai2017}
Z. Dai, P. R. Shepherd, and R. J. Watson, \emph{UAV-aided source localization in urban environments based on ray launching simulation},\hskip 1em plus
  0.5em minus 0.4em\relax IEEE International Symposium on Antennas and Propagation \& USNC/URSI National Radio Science Meeting, 2017.

\bibitem{lwu2020}
L. Wu, D. He, B. Ai, J. Wang, H. Qi, K. Guan, and Z. Zhong, \emph{Artificial Neural Network Based Path Loss Prediction for Wireless Communication Network},\hskip 1em plus
  0.5em minus 0.4em\relax IEEE Access, Nov., 2020.
  
\bibitem{music1986}
M. Kaveh and A. J. Barabell, \emph{The statistical performance of the MUSIC and the minimum-norm algorithms in resolving plane waves in noise},\hskip 1em plus
  0.5em minus 0.4em\relax IEEE Transactions on Acoustics, Speech, and Signal Processing, Vol. 34, No. 2, Apr., 1986.

\bibitem{ESPRIT1997}
K. T. Wong and M. D. Zoltowski, \emph{Uni-vector-sensor ESPRIT for multisource azimuth, elevation, and polarization estimation},\hskip 1em plus
  0.5em minus 0.4em\relax IEEE Transactions on Antennas and Propagation, Vol. 45, Issue 10, 1997.

\bibitem{mpastorino2005}
M. Pastorino and A. Randazzo, \emph{A Smart Antenna System for Direction of Arrival Estimation Based on a Support Vector Regression},\hskip 1em plus
  0.5em minus 0.4em\relax IEEE Transactions on Antennas and Propagation, Vol. 53, No. 7, Jul., 2005.

\bibitem{zliu2018}
Z. Liu, C. Zhang, and P. S. Yu, \emph{Direction-of-Arrival Estimation Based on Deep Neural Networks With Robustness to Array Imperfections},\hskip 1em plus
  0.5em minus 0.4em\relax IEEE Transaction on Antennas and Propagation, Vol. 66, No. 12, Dec., 2018.

\bibitem{myang2020}
M. Yang, B. Ai, R. He, H. Chen, Z. Ma, and Z. Zhong, \emph{Angle-of-Arrival Estimation for Vehicle-to-vehicle Communications based on Machine Learning},\hskip 1em plus
  0.5em minus 0.4em\relax International Conference on Wireless Communications and Signal Processing (WCSP), Dec., 2020.

\bibitem{obialer2019}
O. Bialer, N. Garnett, and T. Tirer, \emph{Performance Advantages of Deep Neural Networks for Angle of Arrival Estimation},\hskip 1em plus
  0.5em minus 0.4em\relax IEEE International Conference on Acoustics, Speech and Signal Processing (ICASSP), 2019.

\bibitem{akhan2019}
A. Khan, S. Wang, and Z. Zhu, \emph{Angle-of-Arrival Estimation Using an Adaptive Machine Learning Framework},\hskip 1em plus
  0.5em minus 0.4em\relax IEEE Communications Letters, Vol. 23, No. 2, Feb., 2019.

\bibitem{jyu2021}
J. Yu, W. W. Howard, D. Tait, and R. M. Buehrer, \emph{Direction-of-Arrival Estimation With A Vector Sensor Using Deep Neural Networks},\hskip 1em plus
  0.5em minus 0.4em\relax IEEE 93rd Vehicular Technology Conference (VTC2021-Spring), 2021.

\bibitem{mgall2020}
M. Gall, M. Gardill, T. Horn, and J. Fuchs, \emph{Spectrum-based Single-Snapshot Super-Resolution Direction-of-Arrival Estimation using Deep Learning},\hskip 1em plus
  0.5em minus 0.4em\relax German Microwave Conference (GeMiC), Mar., 2020.

\bibitem{lwu2019}
L. Wu, Z. Liu, and Z. Huang, \emph{Deep Convolution Network for Direction of Arrival Estimation With Sparse Prior},\hskip 1em plus
  0.5em minus 0.4em\relax IEEE Signal Processing Letters, Vol. 26, No. 11, Nov., 2019.
  
\bibitem{dai2020}
Z. Dai and F. Podd, \emph{A Power-Efficient BLE augmented GNSS Approach to Site-Specific Navigation},\hskip 1em plus
  0.5em minus 0.4em\relax IEEE/ION Position, Location and Navigation Symposium (PLANS), 2020.
  
\bibitem{abdallah2018}
A. A. Abdallah, K. Shamaei, and Z. M. Kassas, \emph{Indoor Positioning Based on LTE Carrier Phase Measurements and an Inertial Measurement Unit},\hskip 1em plus
  0.5em minus 0.4em\relax Proceedings of the 31st International Technical Meeting of the Satellite Division of The Institute of Navigation (ION GNSS+), 2018.
  
\bibitem{zli2018}
Z. Li, T. Braun, X. Zhao, Z. Zhao, F. Hu, and H. Liang, \emph{A Narrow-Band Indoor Positioning System by Fusing Time and Received Signal Strength via Ensemble Learning},\hskip 1em plus
  0.5em minus 0.4em\relax IEEE Access, Vol. 6, Jan., 2018.

\bibitem{rahman2018}
M. T. Rahman, N. Tadayon, S. Han, and S. Valaee, \emph{LocHunt: Angle of Arrival Based Location Estimation in Harsh Multiplath Environments},\hskip 1em plus
  0.5em minus 0.4em\relax IEEE Global Communications Conference (GLOBECOM), Dec., 2018.
  
\bibitem{probyns2018}
P. Robyns, P. Quax, W. Lamotte, and W. Thenaers, \emph{A Multi-Channel Software Decoder for the LoRa Modulation Scheme},\hskip 1em plus
  0.5em minus 0.4em\relax Proceedings of the 3rd International Conference on Internet of Things, Big Data and Security (IoTBDS), 41-51, 2018.

\bibitem{amarquet2020}
A. Marquet, N. Montavont, and G. Papadopoulos, \emph{Towards an SDR implementation of LoRa},\hskip 1em plus
  0.5em minus 0.4em\relax Computer Communications, Vol. 153, Pages 595-605, Mar., 2020.

\bibitem{nbnilam2021}
N. BniLam, D. Joosens, M. Aernouts, J. Steckel, and M. Weyn, \emph{LoRay: AoA Estimation System for Long Range Communication Networks},\hskip 1em plus
  0.5em minus 0.4em\relax  IEEE Transactions on Wireless Communications, Vol. 20, Issue 3, Mar., 2021.

\bibitem{zli2020}
Z. Li, D. Zhang, Q. Zhu, H. Gu, S. Huang, Y. Kuang, and Y. Liu, \emph{Application Research on DOA Estimation Based on SoftwareDefined Radio Receiver},\hskip 1em plus
  0.5em minus 0.4em\relax 2nd International Conference on Electronic Engineering and Informatics, 2020.

\bibitem{hxiang2019}
H. Xiang, B. Chen, M. Yang, T. Yang, and D. Liu, \emph{A Novel Phase Enhancement Method for Low-Angle Estimation Based on Supervised DNN Learning},\hskip 1em plus
  0.5em minus 0.4em\relax IEEE Access, Vol. 7, Jun., 2019.

\bibitem{xhe2010}
X. He, B. Jiang, J. Zhong, Y. Sun, and Z. Liu, \emph{Direction Of Arrival Estimation Based On Smooth Support Vector Regression},\hskip 1em plus
  0.5em minus 0.4em\relax 2nd International Conference on Future Computer and Communication, May., 2010.

\bibitem{wzhu2019}
W. Zhu and M. Zhang, \emph{A Deep Learning Architecture for Broadband DOA Estimation},\hskip 1em plus
  0.5em minus 0.4em\relax IEEE International Conference on Communication Technology, 2019.
  
\bibitem{mcomiter2018}
M. Comiter and H. T. Kung, \emph{Localization Convolutional Neural Networks Using Angle of Arrival Images},\hskip 1em plus
  0.5em minus 0.4em\relax IEEE Global Communications Conference (GLOBECOM), Dec., 2018.

\bibitem{wqzheng2015}
W. Q. Zheng, Y. X. Zou, and C. Ritz, \emph{Spectral mask estimation using deep neural networks for inter-sensor data ratio model based robust DOA estimation},\hskip 1em plus
  0.5em minus 0.4em\relax IEEE International Conference on Acoustics, Speech and Signal Processing (ICASSP), Apr., 2015.

\bibitem{ysun2018}
Y. Sun, J. Chen, C. Yuen, and S. Rahardja, \emph{Indoor Sound Source Localization With Probabilistic Neural Network},\hskip 1em plus
  0.5em minus 0.4em\relax IEEE Transactions on Industrial Electronics, Vol. 65, No. 8, Aug., 2018.
  
\bibitem{mjbianco2020}
M. J. Bianco, S. Gannot, and P. Gerstoft, \emph{Semi-Supervised Source Localization with Deep Generative Modeling},\hskip 1em plus
  0.5em minus 0.4em\relax International Workshop on Machine Learning for Signal Processing, Sep., 2020.
  
\bibitem{sadavanne2018}
S. Adavanne, A. Politis, and T. Virtanen, \emph{Direction of Arrival Estimation for Multiple Sound Sources Using Convolutional Recurrent Neural Network},\hskip 1em plus
  0.5em minus 0.4em\relax 26th European Signal Processing Conference (EUSIPCO), Sep., 2018.
  
\bibitem{yhe2021}
Y. He, N. Trigoni, and A. Markham, \emph{SoundDet: Polyphonic Sound Event Detection and Localization from Raw Waveform},\hskip 1em plus
  0.5em minus 0.4em\relax International Conference on Machine Learning (ICML), Jul., 2021.
  
\bibitem{amelbir2021}
A. M. Elbir and K. V. Mishra, \emph{Sparse Array Selection Across Arbitrary Sensor Geometries With Deep Transfer Learning},\hskip 1em plus
  0.5em minus 0.4em\relax IEEE Transactions on Cognitive Communications and Networking, Vol. 7, No. 1, Mar., 2021.
  
\bibitem{yshi2019}
Y. Shi, K. Davaslioglu, Y. E. Sagduyu, W. C. Headley, M. Fowler, and G. Green, \emph{Deep Learning for RF Signal Classification in Unknown and Dynamic Spectrum Environments},\hskip 1em plus
  0.5em minus 0.4em\relax IEEE International Symposium on Dynamic Spectrum Access Networks (DySPAN), 2019.

\bibitem{arraytrack}
J. Xiong and K. Jamieson, \emph{Arraytrack: A fine-grained indoor location system},\hskip 1em plus
  0.5em minus 0.4em\relax 10th USENIX Symposium on Networked Systems Design and Implementation (NSDI), 2013

\bibitem{wiwear}
V. Tran, A. Misra, J. Xiong, R. Balan,  \emph{Wiwear: Wearable sensing via directional wifi energy harvesting},\hskip 1em plus
  0.5em minus 0.4em\relax IEEE International Conference on Pervasive Computing and Communications (PerCom), 2019

\bibitem{zooghby2000}
A. H. E. Zooghby, C. G. Christodoulou, and M. Georgiopoulos, \emph{A Neural Network-Based Smart Antenna for Multiple Source Tracking},\hskip 1em plus
  0.5em minus 0.4em\relax IEEE Transactions on Antennas and Propagation, Vol. 48, No. 5, May, 2000.
  
\bibitem{malygin2017}
I. V. Malygin, S. A. Belkov, M. R. Usvyatsov, and A. D. Tarasov, \emph{Radio Signal Detection Using Machine-Learning Approach},\hskip 1em plus
  0.5em minus 0.4em\relax Proceedings of Information Technologies, Telecommunications and Control Systems (ITTCS), 2017.
  
\bibitem{mahfoudi2019}
M. N. Mahfoudi, G. Sivadoss, O. B. Korachi, T. Turletti, and W. Dabbous, \emph{Joint range extension and localization for LPWAN},\hskip 1em plus
  0.5em minus 0.4em\relax Internet Technology Letters, Wiley, 2019.
  
\bibitem{capon1995}
P. Stocia, P. Handel, and T. Soderstrom, \emph{Study of Capon method for array signal processing},\hskip 1em plus
  0.5em minus 0.4em\relax Circuits, Systems and Signal Processing, Vol. 14, No. 6, 1995.

\end{thebibliography}
